\documentclass[10pt]{slides}
\usepackage[dvips]{graphics}
\usepackage{color}

\newcommand{\be}{\begin{equation}}
\newcommand{\ee}{\end{equation}}
\newcommand{\bea}{\begin{eqnarray}}
\newcommand{\eea}{\end{eqnarray}}

\begin{document}

\begin{slide}

\begin{flushright}
PHENO 2000\\
{\tiny  April 17 2000}
\end{flushright}

{\huge {Lattice simulations of the strange quark mass and Fritzsch texture}}
\vskip 4in

Biswajoy Brahmachari\\
Indiana University

\end{slide}
\begin{slide}

\small
{
Strange quark mass has been traditionally calculated using
current algebra mass ratio
\bea
&& {{m_s \over {m_u + m_d}}=12.6 \pm 0.5} \label{eq1}
\eea
Equation (\ref{eq1}) is evaluated using  values of ($m_u + m_d$). 
At the two-loop level of perturbative QCD calculations which
include non-perturbative corrections up to dimension six one
has the result
\bea
&& {(m_u + m_d)({\rm~1~GeV})=15.5 \pm 2.0 {\rm~ MeV}} \label{eq2}
\eea
So we obtain
\bea
&& {m_s({\rm 1~GeV}) = 195 \pm 28 } {\rm ~~MeV} \nonumber\\
&& {m_s({\rm 2~GeV})=150 \pm 21 {\rm ~~MeV}} \label{eq3}
\eea

The ratio
\bea
&& {m_s({\rm 2~GeV})/ m_s({\rm 1~GeV})=0.769} \nonumber \\
&& {{\rm for~~} \alpha_s(m_Z)=0.118} \label{eq4}
\eea
can be obtained by solving renormalization group
equations.

A systematic uncertainty in this result
remains in reconstruction of so called { `spectral function'} from
experimental data of resonances. When a different functional form of the
resonance is  adopted, and three loop order perturbative QCD
theory is used one obtains
\bea
&& {(m_u + m_d)~({\rm 1~GeV})=12.0 \pm 2.5 ~{\rm MeV}} \label{eq5}
\eea
}

\end{slide}
\begin{slide}

\small{

With (\ref{eq5}) and (\ref{eq1}) one gets
\bea
&& {m_s({\rm 1 GeV})=151 \pm 32~{\rm MeV}} \nonumber \\
&& {m_s({\rm 2 GeV})=116 \pm 24 ~{\rm MeV}} \label{eq6}
\eea
\marginpar{{\tiny { H.Leutwyler, Phys.Lett. { B378},313 (1996)}}}

It has been remarked by Leutwyler that it is indeed difficult to account for 
{ vacuum fluctuations}, or sea quark effects generated by
quarks of small masses in  perturbative QCD calculations. Thus,
numerical simulations  of strange quark mass on a lattice becomes
rather attractive, especially if the simulation includes virtual light
quark loop effects.

\noindent ${\bullet}$ Up and down type quarks differ only in 
the $U(1)_{em}$ quantum numbers in an effective theory where the gauge 
symmetry is $SU(3)_c \times U(1)_{em}$.

\noindent ${\bullet}$ We are describing the lattice in terms of a 
theory at the scale of a few GeVs where light quark masses are to be described 
in terms of { observables relevant to their own scales}, which are
meson masses and decay constants.

\noindent ${\bullet}$ 
Lattice simulation
determines $m_s$, $m_u + m_d \over 2$ and the lattice
spacing $a$ using three hadronic observables. They
can be chosen, { for example},  $M_\pi, M_{K^*}$, and $f_\pi$.
Scale $a$ can also be taken as a function of some other  
observable, for example, it may be chosen as $a(M_n)$ or $a(M_\Delta)$
etc.
}
\end{slide}
\begin{slide}
\small{
\marginpar{{\tiny {R.Gupta and T.Bhattacharya, Phys.Rev. { D55},
7203(1997)}}}
\noindent ${\bullet}$ 
how do we describe quark masses when the theory
is living on a discritized lattice? { Wilson-like} fermions are, for example
\be
{a~m_{bare}=log~(1+(1/{2 \kappa} - 1/{2 \kappa_c}))}
\label{eq7}
\ee
In the continuum limit we have $a \rightarrow 0$, and there one gets the 
hopping parameter $\kappa=\kappa_c=1/8$.

\noindent ${\bullet}$ 
To compare the result with
experiment, one has to calculate the $\overline{MS}$ mass at a scale 
$\mu$ starting from the lattice estimate of the bare mass using, for example, the 
mass renormalization constant $Z_m(\mu)$ relating the lattice regularization
scheme to the continuum regularization scheme.

\noindent ${\bullet}$ 
Final results of the physical
quark mass for various definitions of the fermion on a lattice differ
$O(a)$ among each other and one expects to get the { same} result of the
physical quark mass in the { continuum limit} when $a \rightarrow 0$.

\noindent ${\bullet}$ 
Beyond the minimal lattice simulation of light quark masses
using the heavy quark effective theory, the next step would
be to incorporate { sea quark effects}. 
They are termed $n_f=2$ { unquenched} lattice simulations.

\noindent ${\bullet}$ 
Bottom quark mass is in the range
\bea
&& {4.1 < m_b(m_b) < 4.4~~{\rm GeV}}
\label{eq8}
\eea
according to review of particle physics (PDG) tables

}

\end{slide}
\begin{slide}

\small{

\begin{center}
\[
\begin{array}{|c|c|c|c|c|}
\hline
ref & quenched & dynamical & spacing& m_s(MeV) \\
\hline
A & yes & & m_\rho & 143 \pm 6 ~~\&~~ 115 \pm 2\\
B & yes & & m_{K^*} & 130 \pm 20 \\
C & yes & & m_{K^*} & 122 \pm 20 \\
D & yes & & m_{K^*} & 111 \pm 12 \\
E & yes & & m_\rho & 110 \pm 31 \\
F & yes & & m_\rho & 108 \pm 4 \\
G & yes & & 1P-1S & 95 \pm 16 \\
\hline
A & & yes & m_\rho & 70~~\&~~80 \\
E & & yes & m_\rho &  68 \pm 19 \\
G & & yes & 1P-1S  & 54-92\\
\hline
\end{array}
\]
\end{center}

\vskip 2cm

\noindent ({{A}}) 
CP-PACS Collaboration (K. Kanaya {\it et al.}), Nucl. Phys. Proc. Suppl. { 73}, 192 (1999).

\noindent ({{B}})
V. Gimenez {\it et al.}, Nucl. Phys. { B540},
472 (1999).

\noindent ({{C}})
C. R. Allton {\it et al.}, Nucl. Phys. {
B489}, 427 (1997).

\noindent ({{D}})
D. Becirevic {\it et al.}, Nucl. Phys. Proc. Suppl. { 73}, 222 (1999).

\noindent ({{E}})
R. Gupta and T. Bhattacharya, Phys. Rev. { D55},
7203 (1997).

\noindent ({{F}})
M. Gockeler {\it et al.}, Nucl. Phys. Proc. Suppl. { 73}, 237 (1999); Phys. Rev.
{ D57}, 5562 (1998).

\noindent ({{G}})
B. J. Gaugh {\it et al.}, Phys. Rev. Lett. {
72}, 1622 (1997).

}
\end{slide}
\begin{slide}

\small{

\marginpar{{\tiny {R.Rattazzi, U.Sarid and  L.J.Hall.
SU-ITP-94-15, May 1994 }}}
Theoretically, one re-expresses the bottom quark mass in terms of
parameters of the Minimal Supersymmetric Standard Model(MSSM). The tree
level contribution which is related straight to the Yukawa texture, and
the one loop contribution due to the dominant gaugino loop can be
accounted individually. Then one can write down the relation
\bea
&& {m_b = m_b^{texture} + m_b^{SUSY}} \nonumber \\
&& {=h_b~{V_F \over \sqrt{2}}~\cos~\beta + {m_b}~{8 \over
3}~g^2_3~{\tan \beta \over 16~\pi^2}~{m_{\tilde{g}}~\mu \over m^2_{eff}}
\label{eq9}}
\eea
Here $m_{\tilde{g}}$ is the gluino mass $\mu$ is the $\mu$ parameter
and $m_{eff}$ is averaged supersymmetry breaking mass scale.

We will discuss  a scenario where the first term of the RHS
of (\ref{eq9}) comes from diaginalizing a Fritzsch Yukawa \\
texture.

\noindent ${\bullet}$ 
Suppose in a two generation case rotation angles of the up and
\marginpar{{\tiny {
 S.Weinberg, 
Trans. NewYork Acad.Sci. { 38}, 185 (1977)
 }}}
the down sectors are $\theta_u$ and $\theta_d$. Then
of the combined quark mixing matrix 
\bea
&& {V=O_u{O_d}^\dagger ~~{\longrightarrow}~~ 
\theta_c=\theta_u \pm \theta_d}
\eea

\noindent ${\bullet}$ 
the ratio of the masses of the first and the second generation satisfies
well the relation
\bea
&& {\tan \theta_c= \sqrt{m_d \over m_s}}.
\label{eq12}
\eea
}
\end{slide}
\begin{slide}
\small{
Fritzsch mass matrices can be thought of as a set of mass matrices which
generalizes (\ref{eq12}) to the following form
\bea
&& {\theta_c=\theta_d \pm \theta_u
~\longrightarrow~\tan^{-1}\sqrt{m_d \over m_s} \pm
\tan^{-1}\sqrt{m_u \over m_c}}
\label{eq13}
\eea
In the three generation case Fritzsch
textures for up and down sectors are given by
\bea
&& {M_U=\pmatrix{0 & a e^{i r} & 0 \cr
             a e^{i r^\prime} & 0 & b e^{ i h} \cr
             0 & b e^{i h^\prime} & c e^{i q} }} \nonumber\\
&& \nonumber\\
&& {M_D=\pmatrix{0 & A e^{i R} & 0 \cr
             A e^{i R^\prime} & 0 & B e^{ i H} \cr
             0 & B e^{i H^\prime} & C e^{i Q} }}
\label{eq14}
\eea
weak mixing matrix is expressed as
\bea
&& {O_U \pmatrix{1 & 0 & 0 \cr
             0 & e^{ i \sigma} & 0 \cr
             0 & 0 & e^{i \tau} } {O_D}^{-1}}
\label{eq16}
\eea
where the phases are
\bea
&& {\sigma =  (r-R)-(h-H)-(h^\prime-H^\prime)+(q-Q)} \nonumber\\
&& {\tau  =  (r-R)-(h^\prime-H^\prime)}
\label{eq15}
\eea
In (\ref{eq16}) $O_U$ and $O_D$ diagonalizes ${M}_U$ and ${M}_D$
in the limit when all the phases vanish.

\noindent { {$\bullet$}}
We will use the approximation of strongly hierarchical eigenvalues.

}
\end{slide}
\begin{slide}

\small{
\[
{
\pmatrix{ 1 
        & - \nu_1 + \mu_1 e^{ i \sigma}
        & \mu_1 ( \nu_2 e^{ i \sigma } - \mu_2 e^{ i \tau }) \cr
           - \mu_1 + \nu_1 e^{ i \sigma}
        & \mu_1 \nu_1 + \mu_2 \nu_2 e^{ i \sigma } + e^{ i \sigma}
        & \nu_2 e^{ i \sigma} - \mu_2 e^{ i \tau}  \cr
           \nu_1 ( \mu_2 e^{ i \sigma } - \nu_2 e^{ i \tau })
        & \mu_2 e^{ i \sigma} - \nu_2 e^{ i \tau}
        & \mu_2 \nu_2 e^{i \sigma} + e^{ i \tau}
          }}
\label{eq22}
\]
where
\bea
\mu_1=\sqrt{m_u \over m_c} ~~~~
\mu_2=\sqrt{m_c \over m_t} ~~~~
\nu_1=\sqrt{m_d \over m_s} ~~~~
\nu_2=\sqrt{m_s \over m_b}
\label{eq21}
\eea

\marginpar{{\tiny {
K.S.Babu and Q.Shafi, Phys.Rev. { D47},5004
(1993)
}}}
\noindent { {$\bullet$}} 
It is easy to check that Fritzsch relations make the top quark mass too light
to be experimentally true ($\sim$ 100 GeV).

\noindent { {$\bullet$}} 
If the flavor symmetries were exact only above the GUT scale, could
a miracle of renormalization group evolution of the masses and mixing
angles make the Fritzsch relations valid at low energy ?

one can choose { before making any other computation}
\bea
\sigma \sim \tau \sim -{\pi \over 2}
\label{eq23}
\eea
Let us set our notations of  mixing angles, the
non-removable phase and eigenvalues of the Yukawa matrices. We adopt the
parameterization
\be
\pmatrix{s_1 s_2 c_3 + c_1 c_2 e^{ i \phi}  & c_1 s_2 c_3 - s_1 c_2 e^{ i
\phi} & s_2 s_3 \cr
s_1 c_2 c_3 - c_1 s_2 e^{ i \phi} & c_1 c_2 c_3 + s_1 s_2 e^{ i \phi} &
c_2 s_3 \cr
-s_1 s_3 & - c_1 s_3 & c_3 }.
\label{eq24}
\ee
}
\end{slide}
\begin{slide}
\tiny{
     In this
parameterization \underbar{eigenvalues} $y_i$ of the Yukawa textures,
three CKM mixing angles and the CP violating phase $\phi$ satisfy
the following renormalization group equations
\marginpar{{\tiny {
S.Naculich, Phys. Rev. { D48}, 5293 (1993)
}}}
\bea
&& 16 \pi^2 {d \over dt} \phi = 0~~, \nonumber \\
&& 16 \pi^2 {d \over dt} \ln \tan \theta_1  = -y^2_t \sin^2 \theta_3~~,
\nonumber\\ 
&&16 \pi^2 {d \over dt} \ln \tan \theta_2  = -y^2_b \sin^2 \theta_3~~,
\nonumber\\
&&16 \pi^2 {d \over dt} \ln \tan \theta_3  = -y^2_t -y^2_b~~,
\nonumber\\
&&16 \pi^2 {d \over dt} \ln y_u =   
-c^u_i g^2_i + 3 y^2_t + y^2_b \cos^2 \theta_2 \sin^2 \theta_3~~,
\nonumber\\
&&16 \pi^2 {d \over dt} \ln y_c =
-c^u_i g^2_i + 3 y^2_t + y^2_b \sin^2 \theta_2 \sin^2 \theta_3~~,
\nonumber\\
&&16 \pi^2 {d \over dt} \ln y_t =
-c^u_i g^2_i + 6 y^2_t + y^2_b cos^2 \theta_3~~,
\nonumber \\
&&16 \pi^2 {d \over dt} \ln y_d =
-c^d_i g^2_i + y^2_t \sin^2 \theta_1 \sin^2 \theta_3 + 3 y^2_b +
y^2_\tau~~,
\nonumber \\
&&16 \pi^2 {d \over dt} \ln y_s =
-c^d_i g^2_i + y^2_t \cos^2 \theta_1 \sin^2 \theta_3 + 3 y^2_b + y^2_\tau
~~,\nonumber \\
&&16 \pi^2 {d \over dt} \ln y_b =
-c^d_i g^2_i + y^2_t \cos^2 \theta_3 + 6 y^2_b + y^2_\tau~~,
\nonumber \\
&&16 \pi^2 {d \over dt} \ln y_e =
-c^e_i g^2_i + 3 y^2_b + y^2_\tau~~,
\nonumber \\
&&16 \pi^2 {d \over dt} \ln y_\mu =
-c^e_i g^2_i + 3 y^2_b + y^2_\tau~~,
\nonumber \\
&&16 \pi^2 {d \over dt} \ln y_\tau =
-c^e_i g^2_i + 3 y^2_b + 4 y^2_\tau~~.
\eea
}
\end{slide}
\begin{slide}
\small{

\begin{center}
\[
\begin{array}{|c|c|c|c|c|c|}
\hline
\alpha_s  & \tan \beta & m_s({\rm 2~GeV})\\
\hline
0.118    &  2  &   59.90~~{\rm~MeV}     \\
0.118    &  10  &  61.52~~{\rm~MeV}     \\
0.118    &  20  &  63.05~~{\rm~MeV}     \\
0.118    &  30  &  66.90~~{\rm~MeV}     \\
\hline
\end{array}
\]
\end{center}

\noindent { {$\bullet$}} 
Implications of results of $n_f=2$ unquenched lattice
simulations of the strange quark mass in the context of the Fritzsch
texture are studied.

\noindent { {$\bullet$}} 
Fritzsch texture demands a large Yukawa
contribution to the bottom quark mass. This is partially canceled
by the supersymmetric loop corrections. 

\noindent { {$\bullet$}} 
We conclude  that 
original Fritzsch texture is consistent with experimental data
if it holds at the GUT scale.

}
\end{slide}
\end{document}